\documentclass{iau}
\usepackage{natbib}
\usepackage{graphicx}
\usepackage{float}

\title[Transformationally decoupling clustering and tracer bias] 
{Transformationally decoupling clustering and tracer bias}

\author[Mark C.\ Neyrinck]   
{Mark C.\ Neyrinck$^1$
\affiliation{$^1$Department of Physics and Astronomy, The Johns Hopkins University, Baltimore, MD 21211 \\ email: {\tt neyrinck@pha.jhu.edu}}}

\newcommand{\apj}{{ApJ}}

\newcommand{\apjl}{{ApJL}}

\newcommand{\mnras}{{MNRAS}}
\newcommand{\nat}{{Nature}}

\newcommand{\pd}{$P_\delta$}

\newcommand{\pg}{$P_{G(\delta)}$}

\pubyear{2014}
\volume{306}  
\jname{Statistical Challenges in 21st Century Cosmology}
\editors{A.F. Heavens, J.-L. Starck, A. Krone-Martins eds.}
\begin{document}

\maketitle

\begin{abstract}
Gaussianizing transformations are used statistically in many non-cosmological fields, but in cosmology, we are only starting to apply them. Here I explain a strategy of analyzing the 1-point function (PDF) of a spatial field, together with the `essential' clustering statistics of the Gaussianized field, which are invariant to a local transformation. In cosmology, if the tracer sampling is sufficient, this achieves two important goals. First, it can greatly multiply the Fisher information, which is negligible on nonlinear scales in the usual $\delta$ statistics. Second, it decouples clustering statistics from a local bias description for tracers such as galaxies.
\keywords{large-scale structure of universe, cosmology: theory, methods: statistical}
\end{abstract}

\firstsection 
\section{Transformations and Information}
Cosmologists have been trained to look at the world through linear two-point statistics: the power spectrum and correlation function of the overdensity field, $\delta=\rho/\bar{\rho}-1$. This is for good reason: linear perturbation theory is naturally expressed in terms of the power spectrum of $\delta$, which sources gravity. The raw power spectrum works well for the nearly Gaussian cosmic microwave background (CMB) as well, and the galaxy and matter density fields on large scales. Also, $\delta$ has the benefit that the power at a given scale is largely invariant if the resolution is increased. But the usual correlation function and power spectrum dramatically lose constraining power in a non-Gaussian field such as the matter or galaxy density field on small scales, so to reach the highest-possible precision in cosmology, other approaches are necessary.

\begin{figure}
  \begin{center}
    \includegraphics[width=\columnwidth]{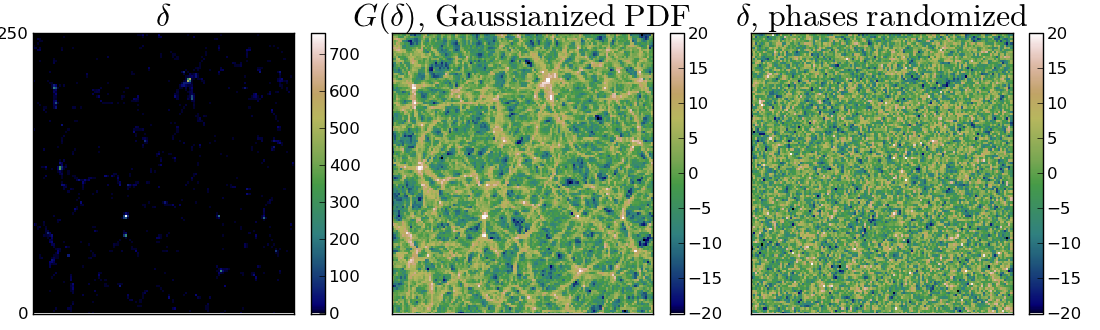}
   
       \includegraphics[width=\columnwidth]{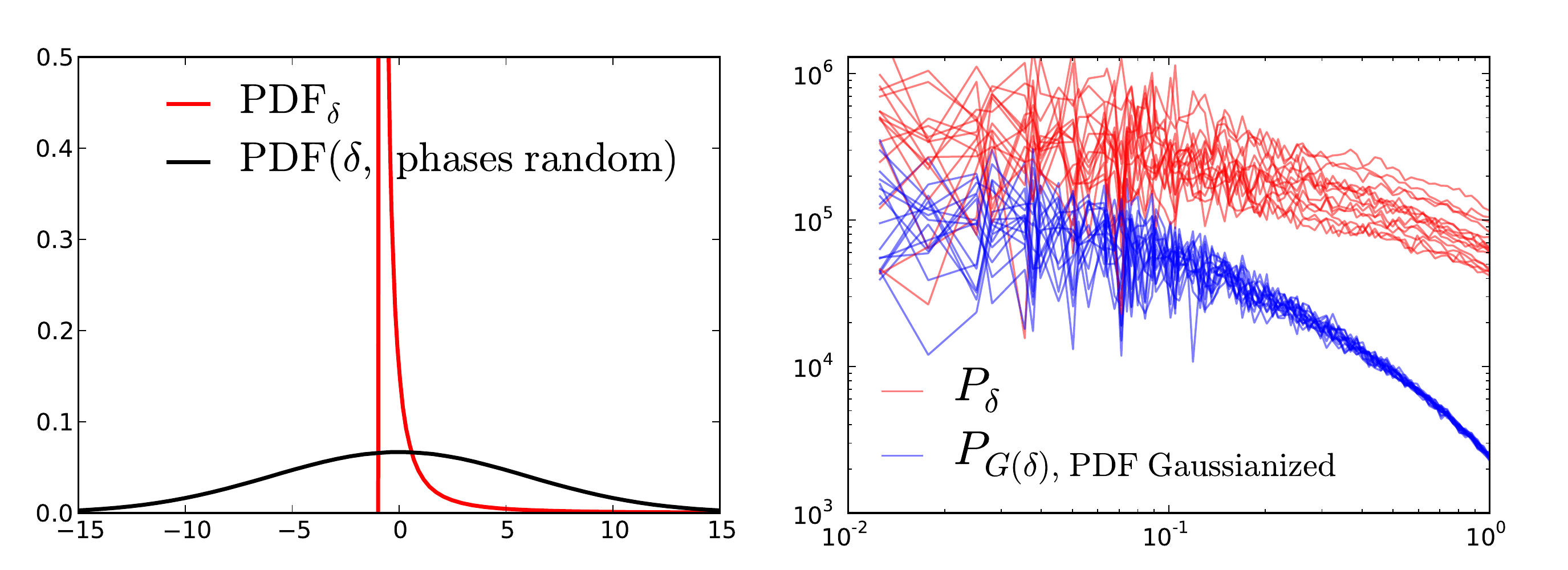}
     \end{center}
  \caption{
  \textit{\textbf{Upper left}}: a quadrant of the the overdensity $\delta=\rho/\bar{\rho} - 1$ in a 2\,$h^{-1}$\,Mpc-thick slice of the 500\,$h^{-1}$\,Mpc Millennium simulation \citep{mill}, viewed with an unfortunate linear color scale.
    \textit{\textbf{Upper middle}}: the same slice, rank-order-Gaussianized, i.e., applying a function on each pixel to give a Gaussian PDF. 
    \textit{\textbf{Upper right}}: the $\delta$ field with its Fourier phases randomized.
    \textit{\textbf{Lower left:}} PDFs of the upper-left and upper-right 3D fields.
    \textit{\textbf{Lower right:}} \pd\ and \pg\ of $\delta$ (red) and $G(\delta)$ (blue), measured from each of several 2D slices such as those shown in the upper-left and middle panels. The wild, coherent fluctuations in \pd\ from slice to slice illustrate its high (co)variance, absent in \pg, which has nearly the covariance properties as in a Gaussian random field.
    }
  \label{fig:showdens}
\end{figure}

Illustrating the nature of non-Gaussianity encountered in cosmology, Fig.\ \ref{fig:showdens} shows a dark-matter density slice from the Millennium simulation, together with the results after two operations: PDF (probability density function) Gaussianization, and randomizing Fourier phases.  The original field and the phase-randomized field look staggeringly different, but have the same $\delta$ two-point statistics. The fact that the power spectrum does not distinguish these fields has been used to demonstrate the need to go to higher-point statistics.  But I assert that this is the wrong way to go, to added complexity and difficulty of analysis.

The main difference by eye is simply in the PDF, shown at bottom left. But even more differences are extractable between the two starkly different fields at `lower order' ($N\le 2$). Applying a nonlinear mapping to a field changes its higher-point statistics.  For example, a Gaussian field, with higher-point ($N>2$) functions identically zero, will sprout nonzero higher-point functions at all orders if a nonlinear function is applied to it \citep{Szalay1988}. I assert that it is useful to define Gaussianized clustering statistics ($[N>1]$-point functions), in which first, the field is PDF-Gaussianized (i.e.\ a mapping is applied to give a field with as Gaussian a 1-point PDF as possible), and then clustering statistics are measured.

The phase randomization already imparts a Gaussian PDF to the upper-right panel, so Gaussianization does nothing. Gaussianization greatly changes the upper-left panel, though, into the upper-middle panel. A mapping was applied to give a Gaussian PDF of variance ${\rm Var}[\ln(1+\delta)]$ for each slice. This changes the (2-point) power spectra in slices from the red to blue curves, as shown at bottom right in Fig.\ \ref{fig:showdens}. This produces a change in both the mean, and, crucially, the covariance of the power spectrum.  The covariance reduction can be seen qualitatively in the lower-right panel of Fig.\ \ref{fig:showdens}, which shows the density power spectra \pd\ and Gaussianized-density power spectra \pg\ for several Millennium Simulation slices.  There are wild fluctuations (i.e.\ variance, or covariance) on small scales of \pd\ that are not present in \pg. These wild fluctuations show up as a drastic reduction in the cosmological-parameter Fisher information in \pd, i.e. an enlargement in error bars \citep[\eg][]{MeiksinWhite1998,rh05,NeyrinckEtal2006,ns07,TakahashiEtal2009}.
Analyzing \pg, on the other hand, enhances cosmological parameter constraints, in principle by a factor of several \citep{Neyrinck2011b}.

My proposal is to measure the 1-point PDF and Gaussianized clustering statistics together; the Gaussianization step not only reduces covariance in the power spectrum itself, but also the covariance between the power spectrum and 1-point PDF. A mathematical reason to analyze the complete 1-point PDF (not simply its moments) is that, if it is sufficiently non-Gaussian, analyzing even arbitrarily high moments does not give all of its information, as has long been known in the statistical literature \citep[\eg][]{aitchison1957lognormal}.  This was pointed out in a cosmological context by \citet{ColesJones1991}, and \citet{Carron2011} more recently emphasized the mistakenness of the conventional wisdom that measuring all $N$-point functions gives all spatial statistical information in cosmology, and the consequences for constraining parameters.

\section{Tracer bias}
From a statistical point of view, measuring the 1-point PDF along with Gaussianized clustering statistics is a superior approach to measuring just the raw power spectrum. One might be fearful of additional irritants from galaxy-vs-matter bias in Gaussianized clustering statistics, but in fact Gaussianization automatically provides a natural framework to incorporate bias issues, potentially {\it simplifying} analysis greatly. Suppose the tracer $\delta_g$, and matter density $\delta_m$ are related by any invertible function.  Then mapping both PDF's onto the same function (for example, a Gaussian) will give precisely the same fields.  This fact has long been exploited in the field of genus statistics \citep[\eg][]{RhoadsEtal1994,GottEtal2009}; any local monotonic density transformation will leave Gaussianized statistics unchanged in a local-bias approximation. Gaussianization was first applied for power spectra in cosmology by \citet{Weinberg1992}. Unfortunately, it does not seem to reconstruct the initial conditions perfectly on small scales, as was the original aim, although it does largely restore the initial power spectrum's shape \citep{NeyrinckEtal2009}. One way to understand this shape restoration is that whereas in \pd, power smears only from large to small scales, power in \pg\ migrates rather evenly both upward and downward in scale \citep{NeyrinckYang2013}. This is because \pd\ is mainly sensitive to overdense regions where fluctuations contract, and a sort of one-halo shot noise appears from sharp spikes \citep{NeyrinckEtal2006}. \pg, on the other hand, is rather equally sensitive to underdense regions as well, where fluctuations expand.

\begin{figure}[H]
  \begin{minipage}[b]{0.6\linewidth}
    \begin{center}
     \includegraphics[width=0.9\columnwidth]{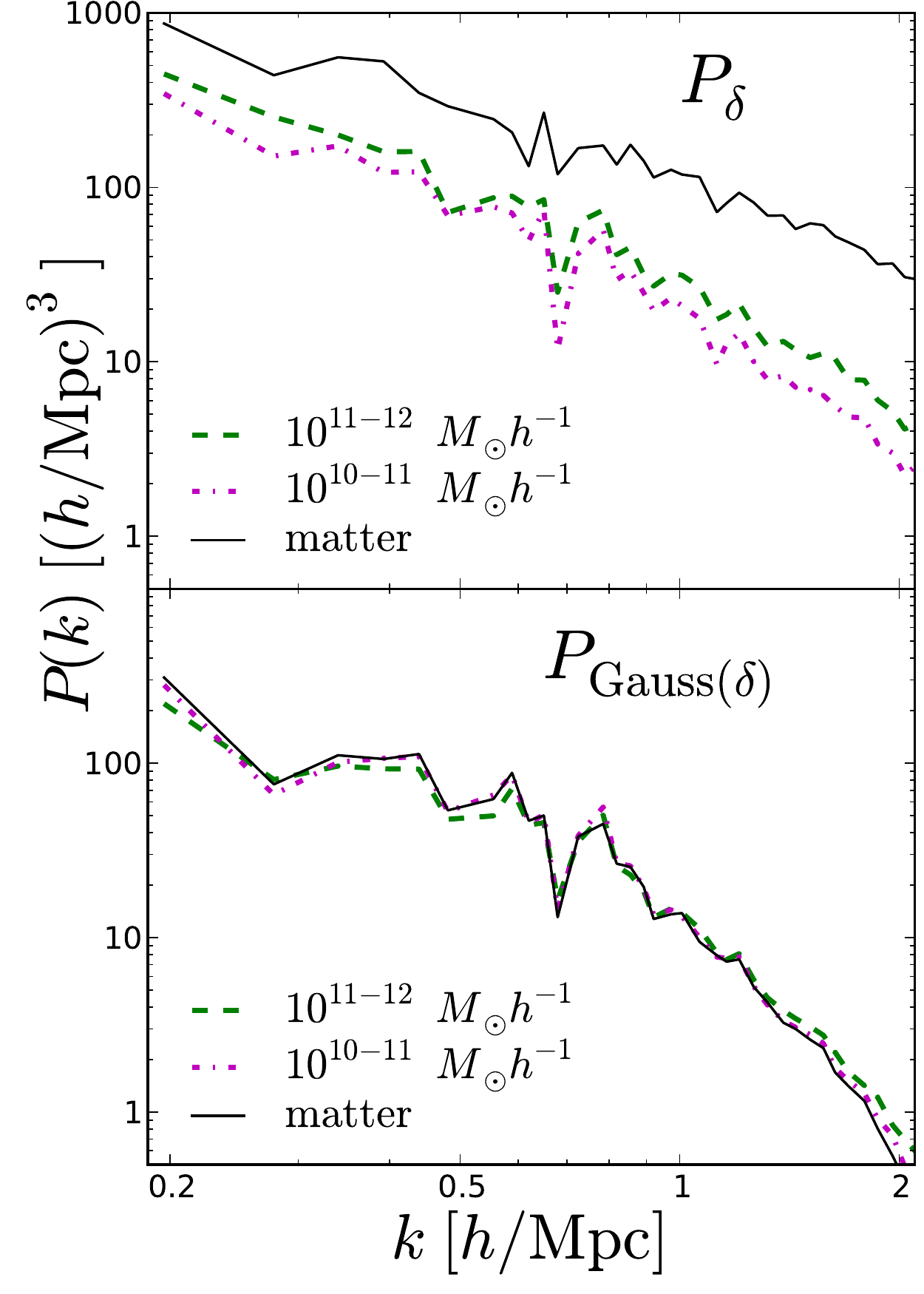}
    \end{center}  
  \end{minipage}
  \begin{minipage}[b]{0.4\linewidth} 
    \begin{center}
      \leavevmode
    \end{center}
    \caption{Power spectra of matter and two mass ranges of haloes in the MIP ensemble-mean fields. The Gaussianized-density power spectra $P_{{\rm Gauss}(\delta)}$ show substantially less difference among the various density fields than the raw density power spectra $P_\delta$, supporting the hypothesis that a local, strictly-increasing density mapping captures the mean relationship between matter and haloes.}
\vskip 2  cm
    \normalsize \hspace{1 em} The usual $\delta$ clustering statistics have large statistical error bars on nonlinear scales, which can swamp errors from sub-optimal measurement. But Gaussianized clustering statistics have great statistical power; with that power comes great responsibility to measure them accurately, which is what we plan to do in future work.
    \label{fig:gausspower}
    \end{minipage}
\end{figure}

Fig.\ \ref{fig:gausspower}, taken from \citet{NeyrinckEtal2014}, shows what Gaussianizing does for different tracers explicitly. It uses the MIP ({\it multim in parvo}) ensemble of $N$-body simulations \citep{AragonCalvo2012}, in which 225 realizations were run with the same initial large-scale modes (with $k<2\pi/(4h^{-1}\,{\rm Mpc})$), but differing small-scale modes. So each simulation gives a different realization of haloes in the same cosmic web.  For Fig.\ \ref{fig:gausspower}, we averaged together the halo and matter density fields over the realizations, and measured the $\delta$ and $G(\delta)$ power spectra.  In the ensemble, there is a rather clean mapping between mean matter density and mean halo density, a power law with an exponential cutoff at low density; see \citet{NeyrinckEtal2014} for details. The correspondence in the Gaussianized power spectra is impressive.

However, in this discussion, we have ignored an important caveat: galaxy discreteness. If empty, zero-density pixels exist, this makes a naive logarithmic transform inapplicable. Also, if there are multiple pixels with the same density, then any assumed mapping from a perfect Gaussian to $\delta$ is not invertible. In this case, $\delta$ can be rank-ordered, and for a $\delta$ appearing multiple times, $G(\delta)$ can be set to the average of all $G(\delta)$ that would map to that range of $\delta$; see \citet{Neyrinck2011b} for more details.  This problem can be negligible, as in the cases of the two figures above, or it can be substantial, in the high-discreteness limit. A rule of thumb is to use pixels large enough to contain on average several galaxies. As long as this scale is in the non-linear regime, it will be fruitful to Gaussianize. A promising new alternative is an optimal transform for a pixelized Poisson-lognormal field \citep{CarronSzapudi2014}. This gives the maximum posterior density from a single pixel in a lognormal-Poisson Bayesian reconstruction, as in \citet{KitauraEtal2010}.

\end{document}